\documentclass[prb,twocolumn,superscriptaddress,showpacs,aps,10pt]{revtex4-1}

\usepackage{color}
\usepackage{amsmath}
\usepackage{graphicx}
\usepackage{amssymb}
\usepackage{amsthm, amscd}
\usepackage{amsfonts}
\usepackage{times}
\usepackage[english]{babel}

\newcommand{\be}{\begin{eqnarray*}}
\newcommand{\ee}{\end{eqnarray*}}
\newcommand{\bk}{{\mathbf{k}}}
\newcommand{\bp}{{\mathbf{p}}}
\newcommand{\bq}{{\mathbf{q}}}
\newcommand{\br}{{\mathbf{r}}}
\newcommand{\bR}{{\mathbf{R}}}
\newcommand{\ba}{{\mathbf{a}}}
\newcommand{\bb}{{\mathbf{b}}}
\newcommand{\bB}{{\mathbf{B}}}
\newcommand{\bA}{{\mathbf{A}}}
\newcommand{\bv}{{\mathbf{v}}}
\newcommand{\bs}{{\mathbf{s}}}
\newcommand{\ket}[1]{\left|{#1}\right\rangle}
\newcommand{\bra}[1]{\left\langle{#1}\right|}

\begin{document}

\title{Comment on ``Elementary formula for the Hall conductivity of interacting systems"}

\author{Steven H. Simon}
\author{Fenner Harper}
\affiliation{Rudolf Peierls Centre for Theoretical Physics, University of Oxford, OX1 3NP, United Kingdom}

\author{N. Read}
\affiliation{Department of Physics, Yale University, P.O. Box 208120, New Haven, CT 06520-8120, USA}

\date{\today}

\begin{abstract}
In a recent paper by Neupert, Santos, Chamon, and Mudry [Phys. Rev. B
  86, 165133 (2012)] it is claimed that there is an elementary formula
for the Hall conductivity $\sigma_{xy}$ of fractional Chern insulators.  We show that the proposed formula cannot  generally be correct, and we  suggest one possible source of the error.  Our reasoning can be generalized 
to show no quantity (such as Hall conductivity) expected to be constant throughout an entire phase of matter can possibly be given as the expectation of any time independent short ranged operator.  
\end{abstract}

\maketitle

\section{The Claim of Neupert et al.}

In Ref.~\onlinecite{Chamon} the following formula for the Hall
conductivity was proposed for gapped two dimensional fractional Chern insulators (FCIs)
\begin{equation}
  \sigma_{xy} =    \frac{2\pi}{{\cal A}}  \sum_{{\bf k} \in\Omega}  F({\bf k}) \langle n({\bf k}) \rangle
  \label{eq:formula}
\end{equation}
where $F({\bf k})$ is the Berry curvature of the occupied energy band, $n({\bf k})=c^\dagger_{\bf
  k} c^{\phantom{\dagger}}_{\bf k}$ is the momentum space density operator, and ${\cal A}$ is the area of the system.   Although it is intended that this formula should hold in the thermodynamic limit, here we have written the formula for a finite system with periodic boundaries so the sum is over all of the discrete allowed $\bf k$ in the Brillouin zone $\Omega$ and the operator $n({\bf k})$ has eigenvalues 0 and 1.   Note also that we measure
$\sigma_{xy}$ in units where $e^2/h=1$, and we have used the conventional normalization of Berry curvature such that the  Hall conductivity for a completely filled Chern band is correctly obtained by this formula. The claim of Ref.~\onlinecite{Chamon} is that Eq.~\ref{eq:formula} should also hold for fractionally filled Chern bands whenever interactions create a FCI ground state with a gap to excitations.   This simple formula has also been invoked as a diagnostic in later work\cite{Neupert2}.   Note that Ref.~\onlinecite{Chamon} also requires that the interactions do not excite electrons out of the single partially filled Chern band in order for Eq.~\ref{eq:formula} to be appropriate.  The generalization to the case of multiple bands is discussed in section \ref{sec:multi} below.   The point of this paper is to show that the claimed formula, Eq.~\ref{eq:formula}  is not generally correct when applied to FCIs.

\section{What is Wrong with the Putative Proof}

\label{sec:putative}

We begin by examining what is wrong with the putative proof of Eq.~\ref{eq:formula} given by the authors of Ref.~\onlinecite{Chamon}.  Although we do not rule out the possibility of additional problems with the putative proof, one particularly obvious shortcoming of their derivation clearly invalidates it.    The argument given in Ref.~\onlinecite{Chamon} relies on reducing the Hall conductivity to an expression (Eq.~3.15 of that work) involving the matrix elements of a many body position operator ${\bf X}$.   While such an operator is well defined in a Hall bar geometry, it is not well defined on a system with periodic boundary conditions ($\bf X$ is only defined modulo the length of the system).     On the other hand, the argument also assumes that the system is completely gapped, and due to edge states, this is not true in Hall bar geometries.   Thus the proof fails for both a system with edges and for a periodic system --- i.e., it fails for any system of any finite size.   Since carefully defined 
thermodynamic limits are always obtained by taking limits of larger and larger finite sized systems, it is inevitable that the proof does not hold for  infinite systems either.

It is conceivable that a similar derivation might be achieved which repairs these particular errors by either using a properly defined operator in place of $\bf X$ or possibly by carefully taking a small $q$ limit of $\sigma_{xy}(q)$.   However, the proof in Ref.~\onlinecite{Chamon} as presented certainly does not do this and currently stands as invalid.  One might wonder if  such a derivation were properly performed, could  the claimed result, Eq.~\ref{eq:formula}, possibly turn out to be correct?   In the remainder of this paper, we will show that this is not possible.

\section{Main Argument}

As mentioned in Ref.~\onlinecite{Chamon},
the Laughlin flux insertion argument demands that the Hall conductivity be
quantized as $\sigma_{xy} = q/N_{gs}$ with $q$ some integer and
$N_{gs}$ the ground state degeneracy.    The way we will show that Eq.~\ref{eq:formula} is not true is by
considering a FCI where $\sigma_{xy}$ is appropriately quantized, then
we will apply a small perturbation which cannot close the gap and
therefore $\sigma_{xy}$ must remain unchanged.  At the same time we
will show that this perturbation must change the value of the above
integral thereby giving a contradiction.

Let us consider a Chern band where the Berry curvature $F({
\bf k})$ is not a constant over the Brillouin zone.   The single particle Hamiltonian $K$ is the kinetic energy within the band.   We  consider an inter-electron interaction $V$ such that at a certain density the ground state of the system is an FCI.   We will assume that the interaction is translationally invariant so that $\bf k$ remains a good quantum number, and we assume $V$ is short-ranged.  (See the appendix for precise definition of ``short-ranged".)  We also assume that the interaction does not mix this Chern band with higher bands; i.e., we assume interactions have been projected to a single band.  Recall that Eq.~\ref{eq:formula} is claimed to hold precisely in this case where the Hamiltonian $H=K+V$ is projected to a single band.    (See section \ref{sec:multi} below for the multi-band case).

In order to keep the system everywhere gapped, and to have momentum $\bf k$ well defined, we will work on a torus geometry.
It is convenient to consider a FCI where the multiple ground states on the torus occur at different values of the momentum\cite{Bernevig}.    Let us consider a
perturbation of the Hamiltonian by a term
$$
             dH = 2 \pi\sum_{{\bf k} \in\Omega}  F({\bf k}) n({\bf k}) $$
which is diagonal in momentum and therefore cannot mix the degenerate ground states or change the Berry curvature.   (Note that
this change in the Hamiltonian is designed to be extensive.  Also note that $dH$ does not introduce long range interactions of any sort, see the discussion in the appendix.)    By   perturbation theory, given a Hamiltonian $ H + \lambda
\, dH $ it is easy to show that as long as the ground state is not an
eigenstate of $dH$, then $\langle dH \rangle$ must decrease for small
positive $\lambda$.  This is quite physical.  If you push on a system
in one place, it responds by moving away from that place.  Thus $
\langle dH \rangle $ (and therefore the right hand side of Eq.~\ref{eq:formula})
must decrease as the perturbation is turned on, providing a
contradiction to the quantization of $\sigma_{xy}$.

There remains only one loophole to this argument.  The statement that
$\langle dH \rangle$ must decrease is true only provided that the
ground state is not itself an eigenstate of $dH$.  (Or more precisely, the above argument fails if the ground state approaches an eigenstate in the thermodynamic limit meaning that any matrix element 
$\langle A| dH|{\rm ground}\rangle$ scales to zero for any ket $|A\rangle$.)   Since $dH$
does not commute with $H$ we certainly do not expect that the ground state
would be an eigenstate of $dH$. Nonetheless, one could still ask how
we know that some unexpected conspiracy does not make this true.

We note that to evade our proof by contradiction, the ground state must also remain an eigenstate of $dH$ (or must approach an eigenstate in the thermodynamic limit) as we slightly deform the original Hamiltonian.   For example, choosing a Hermitian operator $S$ we could construct
\begin{equation}
\label{eq:canonical}
  \tilde H = e^{i \alpha S} H e^{-i \alpha S} = H  + i \alpha [S,H] + \ldots
\end{equation}
with small $\alpha$ and treat the correction terms as a small perturbation to our original Hamiltonian.  By choosing $S$ to be a short-ranged, Hermitian, translationally-invariant operator within our fractionally filled band, we obtain $\tilde H$  having these same properties as well.  (We thank the authors of Ref.~\onlinecite{Chamon} for emphasizing to us the necessity of considering only short-ranged interactions\cite{Chamon2}.)  Since this transformation is canonical, the spectrum of the Hamiltonian remains unchanged.  It is then easy to show that in order for the ground state to remain an eigenstate of $dH$ we must have, order by order,  $[S,dH]$ annihilates the ground state, $[S,[S,dH]]$ annihilates the ground state and so forth.   It seems almost impossible that this should be true for all possible operators $S$.   In section \ref{sec:calc} below we will show even further evidence that this cannot generally be the case.   However, first it is useful to look at a simple example of a Chern band for 
clarity.

\section{A Useful Example}
\label{sec:Hof}

Let us consider the Harper-Hofstadter model for a charged particle hopping on a square lattice of unit lattice constant in the presence of uniform magnetic field, which we choose to provide a flux per
plaquette of $1/N$ with $N$ a large integer.  It turns out that the
lowest band (analogous to the lowest Landau band) has energy and Berry
curvature given by the forms
\begin{eqnarray*}
E({\bf k}) - \bar E &=&  a  (\cos (k_x N)  + \cos ( k_y N) ) \\
F({\bf k}) - \bar F  &=& b (\cos (k_x N)  + \cos (k_y N))
\end{eqnarray*}
where $\bar E$ and $\bar F$ are the average energy and Berry curvature
over the Brillouin zone ($k_x \in [0,2\pi/N]$ and $k_y \in [0,2\pi]$).  Both $a$ and $b$ are constants exponentially small in
$N$, but finite for $N$ not infinite.  The corrections to these
functional forms are smaller by a factor which vanishes quickly in the
large $N$ limit.  These statements, which will be demonstrated in
another work\cite{Fenner}, are easy to check numerically.

Let us write the Hamiltonian as  $H = V + K$ where $V$ is the interaction term and
$$
K = \sum_{{\bf k} \in\Omega}  E({\bf k})  n({\bf k})
$$ is the kinetic term.  We know that for large $N$ this lowest band is
very close to a Landau level.  For appropriately chosen electron
density and appropriately chosen interaction it is clear that we can
produce a FCI ground state.

Because of the proportionality of $E({\bf k}) - \bar E$ with $F({\bf k})-\bar F$ we can rewrite
$
dH = c_1 K + c_2
$
with $c_1$ and $c_2$ constants.  (Note that, at least in this case, $dH$ is a perfectly well behaved short-ranged operator.  See the appendix for an argument that this is true generally.)  If the ground state were an
eigenstate of $dH$ it would also be an eigenstate of $K$ and hence also of $V$, even though $K$ and $V$ do not commute.    This would
imply that the ground state would be unchanged as we change the
relative strengths of the kinetic energy $K$ versus interaction energy $V$, and further that the eigenvalue of $K$ is unchanged as the details of the interaction $V$ are perturbed in any way as well.
Except in trivial cases where the band is either completely filled or completely empty, such a set of coincidences is almost obviously impossible.

\section{Further Argument}
\label{sec:calc}

At this point we have shown that in order for Eq.~\ref{eq:formula} to hold while retaining quantization of $\sigma_{xy}$, the ground state must always be an eigenstate of the operator $dH$.  Further, the ground state must be annihilated by $[S,dH]$ for any translationally-invariant, Hermitian, short-ranged operator $S$.  Our strategy now is to suppose these statements are actually true (unlikely though they may seem) and we will show that we can generate even more unlikely conclusions and finally a contradiction, thus invalidating the original assumption that Eq.~\ref{eq:formula} holds.

Let us consider a two-electron momentum conserving short-ranged interaction entirely within our single fractionally filled Chern band.
(Note that, as discussed in the appendix, the projection of a short ranged operator to a single band is still short ranged.)   Such a general operator can be written as
$$
  S= \frac{1}{\cal A}\sum_{{\bf p},{\bf q},{\bf r},{\bf s} \in \Omega} s({\bf p},{\bf q},{\bf r},{\bf s}) \, c^\dagger_{\bf p}   c^\dagger_{\bf r} c^{\phantom{\dagger}}_{\bf q} c^{\phantom{\dagger}}_{\bf s}
\,\, \delta_{{\bf p}+{\bf r},{\bf q}+{\bf s}}  + h.c.
$$
with $\delta$ a Kronecker delta to enforce momentum conservation.   It is then easy to show that  $U = i [S,dH]$ is the two electron interaction
\begin{equation}
\label{eq:Ueq}
  U = \frac{1}{\cal A}\sum_{{\bf p},{\bf q},{\bf r},{\bf s} \in \Omega} u({\bf p},{\bf q},{\bf r},{\bf s}) \, c^\dagger_{\bf p} c^\dagger_{\bf r}  c^{\phantom{\dagger}}_{\bf q}  c^{\phantom{\dagger}}_{\bf s}
\,\, \delta_{{\bf p}+{\bf r},{\bf q}+{\bf s}}  + h.c.
\end{equation}
where
\begin{equation}
\label{eq:usg}
 u({\bf p},{\bf q},{\bf r},{\bf s}) = i s({\bf p},{\bf q},{\bf r},{\bf s}) \,\,  g({\bf p},{\bf q},{\bf r},{\bf s})
\end{equation}
with  
\begin{equation}
\label{eq:gdef}
g({\bf p},{\bf q},{\bf r},{\bf s})  =
F({\bf p}) - F({\bf q})+ F({\bf r}) - F({\bf s})
\end{equation}
A rough argument is now as follows.  Given almost any short-ranged two-electron interaction $U$ we can find a short-ranged operator $S$ such that $i [S,dH]=U$ and therefore this $U$ must annihilate  the ground state.   We will be precise about why we say ``almost any" in the next paragraph, but for now we note that if we were able to construct any $U$ as a commutator $i [S,dH]$, then any short-ranged interaction $U$ would have to annihilate the ground state, which is an absurd conclusion, and therefore would disprove the original assumption of Eq.~\ref{eq:formula}.   

Now this simple argument, although very suggestive, is not rigorous as it stands, because there are a some interactions $U$ which cannot be constructed as a commutator of a short-ranged $S$.   To see this note that the function $g$ can equal zero for certain values of ${\bf p},{\bf q},{\bf r},{\bf s}$ which means that the set of functions that we should consider is restricted by $u$ being also zero for the same combinations of ${\bf p},{\bf q},{\bf r},{\bf s}$.

To examine this more closely, so long as $F({\bf k})$ is not a constant, in the space of momentum conserving ${\bf p},{\bf q},{\bf r},{\bf s}$ we have  $g=0$ along a submanifold of co-dimension one, or along a set of measure zero among all of the allowed momentum conserving combinations of ${\bf p},{\bf q},{\bf r},{\bf s}$.  (See for example, the explicit Hofstadter case discussed above.)  Thus any function $u$ that vanishes along this submanifold conisistent with Eqs.~\ref{eq:usg} and \ref{eq:gdef}  must give an operator $U$ that annihilates the grounds state.  Given that one can construct an infinitly large variety of such operators, including an infinite variety of short-ranged operators, except in the trivial cases of a completely filled or completely empty band, it seems absurd that the ground state should be annihilated by all of these operators.   (Indeed, one could repeat the argument for ($N > 2$)-body operators as well and generate infinitely more operators which must annihilate the ground state 
too!).

One can go further in making this argument even stronger.  We will argue here that the operators of the form $i [S,dH]$ with short-range $S$ are dense in the space of all short range operators $U$.   What we mean by this is that we can approximate any short-ranged $\tilde U$ with some $U = i [S,dH]$ where $S$ is short ranged, and where matrix elements of $U$ are arbitrarily close to those of $\tilde U$.    Thus, although we cannot precisely construct any short ranged $\tilde U$ as a commutator $i [\tilde S,dH]$ of a short ranged $\tilde S$, we can come arbitrarily close, and, as we will discuss below, this will be enough to justify the above rough argument.

Let us define a function $f({\bf p},{\bf q},{\bf r},{\bf s})$, real analytic in its arguments, which is very close to 1 everywhere except in a small but finite region of (momentum) scale $\xi$  around the submanifold of co-dimension one where $g({\bf p},{\bf q},{\bf r},{\bf s})=0$.   In this region we will let $f$ go to zero on the same submanifold where $g$ is zero, and outside of this region we let $f$ approach 1 pointwise as $\xi$ is taken to zero.    Explicitly we may take $f = 1 - \exp(-\alpha |g|^2  / \xi)$ with $\alpha$ a constant taken to be a typical momentum scale for $g$ divided by the typical magnitude of $g$ squared.   Given some arbitrary short ranged interaction $\tilde U$ we can then define
\begin{eqnarray*}
  u_\xi({\bf p},{\bf q},{\bf r},{\bf s}) &=& f({\bf p},{\bf q},{\bf r},{\bf s}) \,\, \tilde  u({\bf p},{\bf q},{\bf r},{\bf s}) \\
  s_\xi({\bf p},{\bf q},{\bf r},{\bf s}) &=& u_\xi ({\bf p},{\bf q},{\bf r},{\bf s}) / g({\bf p},{\bf q},{\bf r},{\bf s})
\end{eqnarray*}
and here we have choosen $f$ go to zero fast enough when $g$ goes to zero such that $s_\xi$ has no divergences. Hence we have removed the problematic region on the submanifold of codimension one, yet we have arranged that the matrix elements $u_\xi({\bf p},{\bf q},{\bf r},{\bf s}) $ approach those of $\tilde u({\bf p},{\bf q},{\bf r},{\bf s})$ almost everywhere in ${\bf p},{\bf q},{\bf r},{\bf s}$ as $\xi$ goes to zero.   From $u_\xi$ and $s_\xi$ we generate corresponding interactions $U_\xi$ and $S_\xi$ satisfying $U_\xi = i [S_\xi,dH]$.   Note further that if the original $\tilde U$ is a short-ranged operator, then $U_\xi$ and $S_\xi$ will be short ranged with a length scale of $1/\xi$ for small enough $\xi$.

Now since $U_\xi$ and $\tilde U$  differ only on a very small region around a submanifold of measure zero, as we take $\xi$ smaller and smaller the matrix elements of $U_\xi$ should converge to those of $\tilde U$.   To be precise about this convergence we will want to take the thermodynamic limit first such that we can consider arbitrarily small increments in momentum space (and hence we can take $\xi$ smaller and smaller).  For example, let us consider scaled matrix elements such as 
\begin{equation}
\label{eq:Ulim}
 \langle {\rm ground} | \tilde U | {\rm ground} \rangle / {\cal A}
\end{equation}
with ${\cal A}$ the area of the system.  This particular matrix element would be the first order perturbation theory correction the energy density of $\tilde U$, and this should approach a constant independent of
system size in the thermodynamic limit.

When taking the thermodynamic limit, one replaces momentum space sums with integrals, and the contribution to the integrals from the region of width $\xi$ around the submanifold where $g=0$ should become negligible as we take $\xi$ to zero.   Thus,  the value of 
\begin{equation}
\label{eq:UxiA}
  \langle {\rm ground} | U_\xi | {\rm ground} \rangle / {\cal A}
\end{equation}
will approach the value of Eq.~\ref{eq:Ulim} asympotically as $\xi$ is taken to zero.    For the moment let us assume that this claimed convergence is true (we will consider the opposite possibility in the next paragraph).  Then, since $U_\xi = i [S_\xi,dH]$ for a short ranged interaction $S_\xi$ we must have $\tilde U$ annihilating the ground state and so we can conclude that in the thermodynamic limit Eq.~\ref{eq:Ulim} must be zero for any short ranged $\tilde U$.  We can similarly argue that in the thermodynamic limit any (area scaled) matrix element of $U_\xi$ will approach that of $\tilde U$, so in fact we can show that at any order in perturbation theory, the effect of the perturbing interaction $\tilde U$ will have to vanish in the thermodynamic limit (the system is gapped so the energy denominators in perturbation theory cannot cause trouble).   Further still we can use this result to show that under perturbation $\tilde U$ the expectation value of any short-ranged operator must not change in the 
thermodynamic limit.   Such conclusions are clearly absurd and allow us to conclude that the original statement, Eq.~\ref{eq:formula} must be incorrect. 

Finally let us return to more closely examine the above claim that Eq.~\ref{eq:UxiA} converges to Eq.~\ref{eq:Ulim} as $\xi$ is taken to zero.  The only way this convergence can fail is if $\tilde U$ acts as a delta function precisely on the surface where $g=0$.  In this case, even for arbitrarily small $\xi$ one cannot remove the small region of size $\xi$ around the singular point.    Since the function $\tilde U$ is assumed short-ranged such singular behavior could only happen if the ground state itself has some sort of (nontopological) long range order that picks out the wavevectors on the submanifold.   This would then require a new type of long range order to exist in all FCIs.    While we cannot exclude the possibility that certain gapped states of matter do have additional (nontopological) long range order, it seems quite unreasonable that all gapped states of matter in partially filled Chern bands should have this. 

One easy example to examine is the FQHE states, which are fluid, and therefore certainly have no long range order.  Being that FCIs are supposed to be continuously connected to their FQHE counterparts\cite{Qi,Thomas},  it should not be the case that a new long range order can appear once a Landau level is modified to have even an infinitesimally small amount of nonuniform Berry curvature.  

In fact, we can construct examples of FCIs such that there is certainly no such long range order.   Let us begin with a simple Landau level.  If we add a weak periodic potential ${\cal V}({\bf r})$ commensurate with the flux (so one unit cell contains one flux quantum) then the Landau level becomes a Chern band with energy dispersion and nonuniform Berry curvature.  If we add interactions such that the original Landau level displays FQHE (necessarily with no long range order), then the addition of a weak ${\cal V}({\bf r})$ can be treated perturbatively and can scatter by reciprocal lattice vectors, but cannot create nontrivial long range order.

To summarize our argument, we began by showing that in order for Eq.~\ref{eq:formula} to generally hold while retaining quantization of $\sigma_{xy}$, any FCI ground state must be an eigenstate of $dH$ and also must be annihilated by $i [S,dH]$ for any (Hermitian, short-ranged, translationally-invariant) operator $S$.   We then showed that for nonconstant Berry curvature we can design $U=i [S,dH]$ to be arbitrarily close  to any (Hermitian, short-ranged, translationally-invariant) interaction $\tilde U$, which means that either the ground state is annihilated by any such interaction (which is absurd) or there is some sort of long range order which makes the ``arbitrarily close" statement not sufficiently close.   Finally we showed that there exist FCIs without such long range order allowing us to conclude definitively that Eq.~\ref{eq:formula} cannot generally be correct.  

The same reasononing we have used above can clearly be generalized to show that the Hall conductivity (or any quantity that is expected to be constant throughout a given phase of matter) cannot generally be given as the expectation of {\it any} short ranged time independent operator.   The fact that our argument applies so generally was pointed out also by the authors of Ref.~\onlinecite{Chamon} in private communication\cite{Chamon2}. 

\section{Multiple Band Case}
\label{sec:multi}

In Ref.~\onlinecite{Chamon} a formula is also given for the case where interactions mix multiple bands (i.e., the system is no longer projected to a single band).  The generalized claim is (compare Eq.~\ref{eq:formula})
\begin{equation}
  \sigma_{xy} =    \frac{2\pi}{{\cal A}}  \sum_{a,b}\sum_{{\bf k} \in\Omega}  F^{ab}({\bf k}) \langle n_{ab}({\bf k}) \rangle
  \label{eq:formulamultip}
\end{equation}
where $a,b$ label the bands,  $n_{ab}({\bf k}) = c^{\dagger}_{{\bf k},a}c^{\phantom{\dagger}}_{{\bf k},b}$, and $F^{ab}$ is now given by $F^{ab} = \epsilon^{\mu \nu} \partial_{\nu} A_{\mu}^{ab}$ where $\epsilon$ is the antisymmetric tensor, $\mu$ and $\nu$ indicate directions $x$ and $y$ in $\bf k$ space, and $\partial_\nu$ means $\partial/\partial k_\nu$ .   Here $$A_{\mu}^{ab}({\bf k}) = -i \langle u^a_{\bf k} | \partial_\mu | u^b_{\bf k} \rangle$$
with $u_{\bf k}^a$ being the Bloch wavefunctions for band $a$.   In this section we argue that this formula cannot be correct either.   

First, we comment that the putative derivation of Eq.~\ref{eq:formula} in Ref.~\onlinecite{Chamon} is performed by first obtaining Eq.~\ref{eq:formulamultip}, and then restricting occupation to be within a single band (one can imagine making the gap between bands infinite).   Thus establishing that Eq.~\ref{eq:formula} is incorrect should also invalidate Eq.~\ref{eq:formulamultip} as well.   Nonetheless, it is useful to directly examine Eq.~\ref{eq:formulamultip}  to see if there are other, potentially clearer, arguments that it must be invalid.

Here we give a different argument against Eq,~\ref{eq:formulamultip} based on gauge invariance.   We are free to redefine the phases of our Bloch wavefunctions as $u_{\bf k}^a \rightarrow e^{i \phi_a({\bf k})} u_{\bf k}^a$ with arbitrary functions $\phi_a$ and let the corresponding operators $c^\dagger_{{\bf k},a}$ transform analogously via $c^\dagger_{{\bf k},a} \rightarrow e^{i \phi_a({\bf k})} c^\dagger_{{\bf k},a}$.    Under this gauge transformation the Hamiltonian is invariant, and $\langle n_{ab}({\bf k})\rangle$ transforms covariantly by a phase.   However, the expression $ F^{ab}({\bf k}) \langle n_{ab}({\bf k}) \rangle$ is not gauge invariant for $a \neq b$.  Considering a simple case where the $\phi_a$ are single valued functions, under this gauge transformation the right hand side of Eq.~\ref{eq:formulamultip} changes by
\begin{equation}
\label{eq:dsig}
 \delta \sigma_{xy} = \frac{4 \pi}{{\cal A}} \sum_{a < b} \sum_{{\bf k} \in\Omega} \epsilon^{\mu \nu} V_\mu^{ab}({\bf  k})  \partial_\nu [ \phi_a({\bf k}) - \phi_b({\bf k}) ] 
\end{equation}
where
$$
 V_\mu({\bf k}) = {\rm Im}\left[A^{ab}_\mu  \langle n_{ab}({\bf k}) \rangle  \right]. 
$$
For the expression Eq.~\ref{eq:formulamultip} to give a gauge invariant answer, Eq.~\ref{eq:dsig} must vanish for all possible choices of the functions $\phi_a$.   Integrating by parts (and performing a functional derivative), this then requires that for all $a,b,\bf k$ we have 
\begin{equation}
  \label{eq:condition}
   \epsilon^{\mu \nu} \partial_\nu V_\mu^{ab}({\bf  k})  = 0.
\end{equation}
It seems like this would require another conspiracy in order to be true.   

To show that no such conspiracy generally occurs, it is easiest to turn to a very simple explicit example.   We consider the case of a flattened two band model on the honeycomb lattice (the flattened Haldane model) with one filled band and without interactions.   In this case, Eq.~\ref{eq:formulamultip} correctly gives a (integer) quantized Hall conductivity corresponding to the Chern number of the filled band  (the off-diagonal $\langle n_{ab} \rangle$ vanishes).  We then imagine adding weak interaction which in general will mix bands --- for simplicity we choose a nearest neighbor interaction.   We then calculate $\langle n_{ab} \rangle$ perturbatively in the interaction.  At first order in the interaction it is easy to establish by direct calculation that Eq.~\ref{eq:condition} is, as suspected, not satisfied everywhere in the Brillouin zone (details of this calculation are given in the Supplementary Material).  Thus we show that Eq.~\ref{eq:formulamultip} is gauge dependent and therefore cannot be 
correct. 

We note that another approach to disprove Eq.~\ref{eq:formulamultip} is to start with band structure having zero Berry curvature  (and zero $F^{ab}$) everywhere in the Brillouin zone, and introduce a time reversal breaking interaction that makes the ground state a FCI with nonzero Hall conductivity.  It turns out to be possible to do this, as we will show in an upcoming publication\cite{Us}.

\section{Summary}

In summary,we have shown that the  formula Eq.~\ref{eq:formula} proposed in Ref.~\onlinecite{Chamon} (and its multi-band generalization, Eq.~\ref{eq:formulamultip})  cannot hold true in general.  In adddition, in section \ref{sec:putative} we point to one particular weakness of the putative proof given by Ref.~\onlinecite{Chamon}.
It is interesting to note that the argument given here can just as well be used to show that the Hall conductivity (or any quantity which is expected to be constant throughout an entire phase of matter) could not generally be given by the expectation of any single short-ranged operator.

\vspace*{10pt}

{\bf Acknowledgements:}  We are grateful for  multiple useful discussions with the authors
of Ref.~\onlinecite{Chamon} as well as with S.~Ryu,
R.~Roy, T.~S.~Jackson, and N.~R.~Cooper. SHS and FH are supported by EPSRC grants  EP/I032487/1
and EP/I031014/1.   NR is supported by NSF Grant
No.~DMR-1005895.   We thank the Simons Center for Geometry and Physics at SUNY Stony Brook
and the Aspen Center for Physics for their hospitality.

\vspace*{10pt}

{\bf  Appendix: Short-Ranged Operators:}   Let $\psi^\dagger({\bf r})$  be a creation operator for an electron at position $\bf r$, where this operator is not intended to be projected to a single band.    We define a one body operator $O$ to be short ranged if matrix elements of the form
$$
\langle 0 | \psi({\bf r_2}) \, O \,  \psi^{{\dagger}}({\bf r_1})|0\rangle
$$
decay exponentially in  $x=|{\bf r_1} - {\bf r_2}|$, where $|0\rangle$ is the vacuum state with no electrons.   When we say the matrix element decays exponentially we mean that the absolute value of the  matrix element is less than $C e^{-x/R}$ for sufficiently large $x$ and for some finite constants $C$ and $R$.  Similarly, a two body interaction $O$ is considered short ranged if matrix elements of the form
$$
\langle 0 | \psi({\bf r_3})\psi({\bf r_4})  \, O \, \psi^\dagger({\bf r_1})\psi^\dagger({\bf r_2}) |0\rangle
$$
decay exponentially in the parameter $x = \min[ \max(|{\bf r_1} - {\bf r_3}|,|{\bf r_2} - {\bf r_4}|),  \max(|{\bf r_1} - {\bf r_4}|,|{\bf r_2} - {\bf r_3}|)]$ .  I.e., it must decay as the positions of the annihilation operators are moved away from the positions of the creation operator, but allowing for the fact that the particles are indistinguishable.   (The more precise definition of exponential decay is as mentioned above.)  We can use similar definitions of ``short ranged'' for $n > 2$ body operators.

A property of a Chern band is that there is no complete orthogonal (Wannier) basis for the band in which all of the basis states are exponentially localized in real space\cite{Wannier}.   Nonetheless, the operator that projects to a single such band is short-range
provided the band does not touch or cross another band.  Therefore projecting a short-ranged operator (such as a generic short ranged $n$-body interaction) to a single band will keep it short ranged. 

Our above argument requires that the operator $dH$ should be a short-ranged single-body operator.   This is obviously true  in the Hofstadter case discussed in section \ref{sec:Hof} above since $dH$ is equivalent to the kinetic energy which is just a short-ranged hopping model.  However, we claim that in fact $dH$ should always be short ranged for any Chern band resulting from a short ranged hopping Hamiltonian.  To see this we realize that the eigenstates $u_{\bf k}^a$ in the Chern band are the solution to a matrix eigenvalue problem with a continuous parameter $\bf k$. Thus (so long as the gap between bands does not close) the eigenstates can be chosen to be real analytic functions of the parameter ${\bf k}$ at least locally over contractible regions in the Brillouin zone, and hence $F({\bf k})$ is real analytic in its argument.  Now since we are discussing a Chern band, we will have to describe different parts of the Brillouin zone in different gauges, but $F$ is gauge invariant so it is everywhere real 
analytic.   Now given some real analytic $F({\bf k})$ periodic in the Brillouin zone, 
we can always Fourier decompose $F$, and reverse engineer a hopping model that reconstructs this $F$.  Since $F$ is  real analytic, the corresponding hopping model must decay exponentially, and projection to a single band does not change the fact that it is short ranged. 

When the single band has non-zero Chern number, the argument that $S_\xi$ is short ranged is a bit subtle since the entire Brillouin zone cannot be described in the same gauge (in the case of zero Chern number, there is no such complication).  Since one must describe ${\bf k}$-space in patches, one might worry whether the discontinuities in $\tilde u$ may cause problems.  However, once we invert the Fourier transform, the discontinuities in $\tilde u$ cancel and we recover a short ranged function $\tilde U$ in real space.  Since $f/g$ is real analytic everywhere in the Brillouin zone, the same remains true when we construct $s_\xi$ from $\tilde u$, hence we obtain a short ranged $S_\xi$.



\newpage

\begin{widetext}

\newpage

\section*{Supplementary Material: Lack of Gauge Invariance in a Particular Model}

We believe that the expression for the Hall conductivity (Neupert Eq.~5.11) is not gauge invariant when a different $\bk$-dependent phase change is applied to each tight-binding band. Below, we outline where this gauge dependence may arise, before giving an example of a system that is explicitly not gauge invariant.  
\subsection{Outline of Argument}
We consider a two-band model (such as the Haldane model), whose wavefunctions may be written using LCAO as 
\be
\psi^\pm_\bk(\br)&=&\sum_{\mathbf{R}}\left[A^\pm(\bk)\phi_A\left(\mathbf{r}-\mathbf{R}\right)+B^\pm(\bk)\phi_B\left(\mathbf{r}-\mathbf{R}+\mathbf{a}\right)\right]e^{i\bk\cdot\mathbf{R}}.
\ee
Here, $\pm$ indicates the band index, subscript $\{A,B\}$ gives the orbital index, $\{A^\pm(\bk),B^\pm(\bk)\}$ are $\bk$-dependent coefficients and the vector $\ba$ gives the displacement between sites within a unit cell. The sum over $\bR$ is over all lattice vectors, and we will eventually set $\ba=0$ to simplify the calculation. The repeating Bloch functions are then
\begin{eqnarray}
u_\bk^\pm(\br)&=&e^{-i\bk\cdot\br}\psi_\bk^\pm(\br)\nonumber\\
&=&\sum_{\mathbf{R}}\left[A^\pm(\bk)\phi_A\left(\mathbf{r}-\mathbf{R}\right)+B^\pm(\bk)\phi_B\left(\mathbf{r}-\mathbf{R}+\mathbf{a}\right)\right]e^{i\bk\cdot(\mathbf{R}-\br)}.\label{bloch}
\end{eqnarray}
 The Schr\"{o}dinger equation may be written 
\be
\left(\begin{array}{cc}
H_0(\bk)+H_z(\bk) & H_x(\bk)-iH_y(\bk) \\
H_x(\bk)+iH_y(\bk) & H_0(\bk)-H_z(\bk)
\end{array}\right)\left(\begin{array}{c}
A^\pm(\bk)\\
B^\pm(\bk)
\end{array}\right)=E^\pm(\bk)\left(\begin{array}{c}
A^\pm(\bk)\\
B^\pm(\bk)
\end{array}\right),
\ee
with $\{H_0(\bk),\mathbf{H}(\bk)\}$ components of the single-particle (e.g. Haldane) Hamiltonian. We solve this to find the coefficients for the two non-interacting bands $\{A^\pm(\bk),B^\pm(\bk)\}$, and since the bands are orthogonal we must have
\be
\left(A^\pm\right)^*\left(A^\mp\right)+\left(B^\pm\right)^*\left(B^\mp\right)=0
\ee
and
\be
\left\langle u_\bk^\pm\right.\ket{u_\bk^\mp}&=&0.
\ee
The band creation operators may be written in the orbital basis as
\begin{equation}
c^\dagger_{\pm,\bk}=A^{\pm}(\bk)c^\dagger_{A,\bk}+B^{\pm}(\bk)c^\dagger_{B,\bk},\label{bandcreationop}
\end{equation}
where the phase of the original orbitals $\phi_{\{A,B\}}(\br)$ is assumed to have been fixed previously. We note that the band creation and annihilation operators directly involve the coefficients $\{A^\pm(\bk),B^\pm(\bk)\}$.

A component of the Berry connection is defined (according to Neupert Eq. 5.4) by
\begin{eqnarray}
\mathbf{A}^{ab}&=&-i\bra{u_\bk^a}\partial_\bk\ket{u_\bk^b}.\label{connection}
\end{eqnarray}
This is just the ordinary (Abelian) Berry connection when $a=b$, but it transforms differently for $a\neq b$. Only the relative phase between $A$ and $B$ is fixed by Schr\"{o}dinger's equation, and so we are free to make a $\mathbf{k}$-dependent gauge transformation that is different for each band. If we do this, the wavefunctions transform as
\be
\ket{u_\bk^\pm}&\to& e^{i\phi_\pm(\bk)}\ket{u_\bk^\pm}\\
\{A^\pm(\bk),B^\pm(\bk)\}&\to& e^{i\phi_\pm(\bk)}\{A^\pm(\bk),B^\pm(\bk)\}.
\ee
and correspondingly 
\begin{equation}
  c^\dagger_{\pm,\bk} \to e^{i \phi_\pm(\bk)}  c^\dagger_{\pm,\bk} \label{ctransform}
\end{equation}
The (diagonalised) Hamiltonian matrix is transformed by the corresponding unitary matrix ${H\to U H U^\dagger}$ with ${U^\dagger(\bk)=\mathrm{diag}\{e^{-i\phi_+(\bk)},e^{-i\phi_-(\bk)}\}}$. According to Eq.~\ref{connection} an off-diagonal Berry connection transforms under this gauge change to
\be
\tilde{\mathbf{A}}^{+-}&=&-i\left[ \bra{u_\bk^+}e^{-i\phi_+}\right]\partial_\bk\left[e^{i\phi_-}\ket{u_\bk^-}\right]\\
&=&e^{i(\phi_--\phi_+)}\bigg[-i\bra{u_\bk^+}\partial_\bk\ket{u_\bk^-}+\partial_\bk\phi_-\left\langle u_\bk^+\right.\ket{u_\bk^-}\bigg]\\
&=&e^{i(\phi_--\phi_+)}\mathbf{A}^{+-},
\ee
where the final term in the second line (proportional to $\partial_\bk\phi_-$) has vanished because the bands are orthogonal. In this way, $\mathbf{A}^{+-}$ does not transform like a `normal' connection $\mathbf{A}\to e^{i\omega}\mathbf{A}+\partial\omega$ under a gauge transformation: it only picks up a phase. The corresponding curvature,
\begin{eqnarray}
F^{+-}&=&\partial_xA_y^{+-}-\partial_yA_x^{+-}\label{curv},
\end{eqnarray}
is not gauge-invariant and under a gauge transformation becomes
\begin{eqnarray}
\tilde{F}^{+-}&=&\partial_x\tilde{A}_y^{+-}-\partial_y\tilde{A}_x^{+-}\nonumber\\
&=&e^{i(\phi_--\phi_+)}\bigg[\partial_xA_y^{+-}-\partial_yA_x^{+-}+i\bigg(A^{+-}_y\partial_x(\phi_--\phi_+)-A^{+-}_x\partial_y(\phi_--\phi_+)\bigg)\bigg].\label{curvtrans}
\end{eqnarray}
We therefore have 
\be
F\to e^{i\omega}F+ie^{i\omega}\bigg[\boldsymbol{\partial}\omega\times\mathbf{A}\bigg]
\ee
for the off-diagonal terms, instead of just
\be
F\to e^{i\omega}F.
\ee
The expression we are most interested in is the Hall conductivity integral (Neupert Eq. 5.11),
\begin{eqnarray}
\sigma_{\mathrm{NSCM}}=\int \mathrm{d}^2\mathbf{k} \,F^{ab}(\bk) \bar{n}^{ab}_\bk\label{hall}
\end{eqnarray}
where the integral is over the Brillouin zone. The occupation number is the expectation
\be
\bar{n}^{ab}_\bk&=&\langle c^\dagger_{a,\bk}c^{\phantom{\dagger}}_{b,\bk}\rangle
\ee
over the ground state of the system. From the definition of the band operators in Eq.~\ref{bandcreationop} and their transformation (Eq.~\ref{ctransform}), we note that the off-diagonal terms change under the $\bk$-dependent gauge transformation according to
\be
\langle c^\dagger_{+,\bk}c^{\phantom{\dagger}}_{-,\bk}\rangle\to e^{i(\phi_+-\phi_-)}\langle c^\dagger_{+,\bk}c^{\phantom{\dagger}}_{-,\bk}\rangle.
\ee
This expectation value therefore gains a phase from the coefficients $\{A^\pm(\bk),B^\pm(\bk)\}$ that compensates for the phase picked up in the Berry curvature (Eq.~\ref{curvtrans}). However, there remains an additive (derivative) term in the curvature that is not compensated for.

The off-diagonal terms in the Hall conductivity integral in the `original' gauge read
\be
\int \mathrm{d}^2\mathbf{k} \,F^{+-}(\bk) \bar{n}^{+-}_\bk&=&\int\left(\partial_xA_y^{+-}-\partial_yA_x^{+-}\right)\langle c^\dagger_{+,\bk}c_{-,\bk}\rangle\, \mathrm{d}^2\mathbf{k}\\
\int \mathrm{d}^2\mathbf{k} \,F^{-+}(\bk) \bar{n}^{-+}_\bk&=&\int\left(\partial_xA_y^{-+}-\partial_yA_x^{-+}\right)\langle c^\dagger_{-,\bk}c_{+,\bk}\rangle\, \mathrm{d}^2\mathbf{k},
\ee
whilst in the transformed gauge they read
\be
\int \mathrm{d}^2\mathbf{k} \,\tilde{F}^{+-}(\bk)\widetilde{ \bar{n}^{+-}_\bk}&=&\int\left(\partial_xA_y^{+-}-\partial_yA_x^{+-}\right)\langle c^\dagger_{+,\bk}c_{-,\bk}\rangle\, \mathrm{d}^2\mathbf{k}\\
&&+i\int\bigg(A^{+-}_y\partial_x(\phi_--\phi_+)-A^{+-}_x\partial_y(\phi_--\phi_+)\bigg)\langle c^\dagger_{+,\bk}c_{-,\bk}\rangle\, \mathrm{d}^2\mathbf{k}\\
\int \mathrm{d}^2\mathbf{k} \,\tilde{F}^{-+}(\bk)\widetilde{ \bar{n}^{-+}_\bk}&=&\int\left(\partial_xA_y^{-+}-\partial_yA_x^{-+}\right)\langle c^\dagger_{-,\bk}c_{+,\bk}\rangle\, \mathrm{d}^2\mathbf{k}\\
&&+i\int\bigg(A^{-+}_y\partial_x(\phi_+-\phi_-)-A^{-+}_x\partial_y(\phi_+-\phi_-)\bigg)\langle c^\dagger_{-,\bk}c_{+,\bk}\rangle\, \mathrm{d}^2\mathbf{k}.
\ee
(The diagonal terms in the Hall conductivity integral \emph{are} gauge invariant).

If we sum over all four contributions to the Hall conductivity integral we find in the original gauge
\be
\sigma_{\mathrm{NSCM}}&=&\int\mathrm{d}^2\bk\,\left[\bar{n}_{\bk}^{+-}\left(\nabla\times\mathbf{A}^{+-}\right)+\bar{n}_{\bk}^{-+}\left(\nabla\times\mathbf{A}^{-+}\right)+\bar{n}_{\bk}^{++}\left(\nabla\times\mathbf{A}^{++}\right)+\bar{n}_{\bk}^{--}\left(\nabla\times\mathbf{A}^{--}\right)\right]
\ee
and in the transformed gauge
\be
\tilde{\sigma}_{\mathrm{NSCM}}&=&\sigma_{\mathrm{NSCM}}+i\int\bigg(A^{+-}_y\partial_x(\phi_--\phi_+)-A^{+-}_x\partial_y(\phi_--\phi_+)\bigg)\langle c^\dagger_{+,\bk}c_{-,\bk}\rangle\, \mathrm{d}^2\mathbf{k}\\
&&+i\int\bigg(A^{-+}_y\partial_x(\phi_+-\phi_-)-A^{-+}_x\partial_y(\phi_+-\phi_-)\bigg)\langle c^\dagger_{-,\bk}c_{+,\bk}\rangle\, \mathrm{d}^2\mathbf{k}.
\ee
If this last term does not vanish, then the expression for the Hall conductivity is not gauge invariant. 

From the definition of the Berry connection (Eq.~\ref{connection}) we see that
\be
\mathbf{A}^{ab}&=&\left[\mathbf{A}^{ba}\right]^*\\
F^{ab}&=&\left[F^{ba}\right]^*\\
\ee
and we also note that
\be
\langle c^\dagger_{+,\bk}c_{-,\bk}\rangle&=&\langle c^\dagger_{-,\bk}c_{+,\bk}\rangle^*.
\ee
We therefore have
\begin{eqnarray}
\tilde{\sigma}_{\mathrm{NSCM}}&=&\sigma_{\mathrm{NSCM}}+i\int\bigg\{\partial_x(\phi_--\phi_+)\left[A^{+-}_y\langle c^\dagger_{+,\bk}c_{-,\bk}\rangle-\left[A^{+-}_y\langle c^\dagger_{+,\bk}c_{-,\bk}\rangle\right]^*\right]\nonumber\\
&&-\partial_y(\phi_--\phi_+)\left[A^{+-}_x\langle c^\dagger_{+,\bk}c_{-,\bk}\rangle-\left[A^{+-}_x\langle c^\dagger_{+,\bk}c_{-,\bk}\rangle\right]^*\right]\bigg\}\, \mathrm{d}^2\mathbf{k}\nonumber\\
&=&\sigma_{\mathrm{NSCM}}-2\int\bigg\{\partial_x(\phi_--\phi_+)\mathrm{Im}\left[A^{+-}_y\langle c^\dagger_{+,\bk}c_{-,\bk}\rangle\right]-\partial_y(\phi_--\phi_+)\mathrm{Im}\left[A^{+-}_x\langle c^\dagger_{+,\bk}c_{-,\bk}\rangle\right]\bigg\}\, \mathrm{d}^2\mathbf{k}.\nonumber\\
\label{gaugeterms}
\end{eqnarray}
In general the off-diagonal expressions $\mathbf{A}^{\pm}\langle c^\dagger_{+,\bk}c^{\phantom{\dagger}}_{-,\bk}\rangle$ will have imaginary parts, and so this final term will be non-zero. We will attempt to find a simple example system for which this final term does not vanish.

To proceed, we note that this extra gauge-dependent term may be written
\begin{eqnarray}
 \int \mathrm{d}^2\bk     \left[  (\partial_{x} \omega)    v_y  - (\partial_y \omega) v_x \right] \equiv \int \mathrm{d}^2\,\bk     \left[(\nabla\omega)\times\mathbf{v} \right] \label{extraterm}
\end{eqnarray}
where we have defined the gauge  transformation function
$$
  \omega = \phi_- - \phi_+,
$$
which is completely arbitrary, and
$$
    {\bf v} = {\rm Im} \left[  {\bf A} ^{-+}  \langle c_{-,\bk}^\dagger c_{+,\bk} \rangle \right] \equiv- {\rm Im} \left[  {\bf A} ^{+-}  \langle c_{+,\bk}^\dagger c_{-,\bk} \rangle \right]. 
$$
Note that partial derivatives are in the $k_x$ and $k_y$ directions, as are the components of the vector $\mathbf{v}$. 
[N.B. This expression for $\bv$ is gauge invariant because it is constructed from an off-diagonal connection $\bA^{-+}$ and an off-diagonal occupation number $\langle c_{-,\bk}^\dagger c_{+,\bk} \rangle$, which gain opposite phases under a gauge transformation. A vector $\bv$ constructed from diagonal components $\bA^{++}$ and $\langle c_{+,\bk}^\dagger c_{+,\bk} \rangle$ would \emph{not} be gauge invariant because $\bA^{++}$ would transform as a $U(1)$ vector potential.]

A general gauge transformation $\omega(\bk)$ will consist of a multi-valued part and a single-valued part. We may write
\be
\omega(\bk)&=&\eta(\bk)+\chi(\bk)
\ee
with $\eta\left(\bk+\bB_i\right)=\eta(\bk)+2\pi n_i$ and $\chi\left(\bk+\bB_i\right)=\chi(\bk)$, and where $\bB_i$ are reciprocal lattice vectors and $n_i$ are (winding) integers. If we further define the real space lattice vectors $\bb_i$, we can write the multi-valued contribution as
\be
\eta(\bk)&=&n_1\bb_1\cdot\bk+n_2\bb_2\cdot\bk
\ee
and transfer all other $\bk$-dependence to the single-valued term $\chi(\bk)$. We then have
\be
\nabla\omega&=&n_1\bb_1+n_2\bb_2+\nabla\chi
\ee
and
\be
\int\mathrm{d}^2\bk\left[\left(\nabla\omega\right)\times\bv\right]&=&\int\mathrm{d}^2\bk\left[\left(n_1\bb_1+n_2\bb_2\right)\times\bv\right]+\int\mathrm{d}^2\bk\left[\left(\nabla\chi\right)\times\bv\right].
\ee
For the Hall conductivity expression to be gauge invariant, the right hand side must vanish. Additionally, for an arbitrary gauge transformation we are free to choose the single-valued component $\chi(\bk)$ and the integers $n_i$ separately, and so each integral on the right hand side must vanish independently. 

For the case we consider below, the first integral vanishes due to the symmetry of the gauge-independent vector $\bv$, and this may hold in general. Even if this is not the case, $\chi(\bk)$ remains arbitrary, and we will show that this means that the second term does not in general vanish.

By integrating by parts, the second integral can be rewritten as
$$
 \int \mathrm{d}^2\bk   \,  \chi    \,  (\nabla \times {\bf v})_z
$$
Since $\chi$ is arbitrary, the only way this integral can be zero is if the curl of $\bf v$ is everywhere zero,
$$
   (\nabla \times {\bf v})_z =  \epsilon^{ij} \partial_i v_j = 0.	
$$
In the next section we will find a system for which this does not hold.
\subsection{Haldane Model with Weak Nearest-Neighbour Interaction}
The model we will consider is the Haldane honeycomb model with its lowest band initially completely filled and the upper band initially completely empty. We will then perturb weakly about this state with a nearest-neighbour interaction so that the off-diagonal expectation values $\langle c_{\pm,\bk}^\dagger c^{\phantom{\dagger}}_{\mp,\bk} \rangle$ become non-zero.
\subsubsection{Single-particle Properties}
To begin, we define some notation and recall the single-particle properties of the Haldane model. The full Haldane Hamiltonian is (using his own notation),
\be
\hat{H}(\bk)&=&2t_2\cos\phi\left[\sum_i\cos(\bk\cdot \bb_i)\right]\mathbf{I}+t_1\left[\sum_i\left[\cos(\bk\cdot \ba_i)\boldsymbol{\sigma}_1+\sin(\bk\cdot \ba_i)\boldsymbol{\sigma}_2\right]\right]\\
&&+\left[M-2t_2\sin\phi\left[\sum_i\sin(\bk\cdot\bb_i)\right]\right]\boldsymbol{\sigma}_3\\
&\equiv&H_0(\bk)\mathbf{I}+H_x(\bk)\boldsymbol{\sigma}_1+H_y(\bk)\boldsymbol{\sigma}_2+H_z(\bk)\boldsymbol{\sigma}_1\\
&\equiv&\left(\begin{array}{cc}
H_0(\bk)+H_z(\bk) & H_x(\bk)-iH_y(\bk) \\
H_x(\bk)+iH_y(\bk) & H_0(\bk)-H_z(\bk)
\end{array}\right)
\ee
where we identify
\be
H_0(\bk)&=&2t_2\cos\phi\sum_i\cos(\bk_i\cdot\bb_i)\\
H_x(\bk)&=&t_1\sum_i\cos(\bk\cdot \ba_i)\\
H_y(\bk)&=&t_1\sum_i\sin(\bk\cdot \ba_i)\\
H_z(\bk)&=&M-2t_2\sin\phi\sum_i\sin(\bk\cdot\bb_i).
\ee
On the ordinary honeycomb lattice we define the lattice vectors
\be
\bb_1=\sqrt{3}a(-1/2,\sqrt{3}/2)\\
\bb_2=\sqrt{3}a(-1/2,-\sqrt{3}/2)\\
\bb_3=\sqrt{3}a(1,0)\\
\ee
and the nearest neighbour displacement vectors
\[
\begin{array}{ccccc}
\ba_1&=&a(\sqrt{3}/2,1/2)&\equiv&\ba-\bb_2\\
\ba_2&=&a(-\sqrt{3}/2,1/2)&\equiv&\ba+\bb_1\\
\ba_3&=&a(0,-1)&\equiv&\ba\\
\end{array}
\]
where $a$ is the side length of a hexagon and $\ba$ is the sublattice displacement vector discussed earlier (that we will eventually set to zero).

To simplify notation, we implicitly define the spherical polar coordinates
\be
H_x(\bk)&=&\left|\mathbf{H}(\bk)\right|\sin\theta_\bk\cos\phi_\bk\\
H_y(\bk)&=&\left|\mathbf{H}(\bk)\right|\sin\theta_\bk\sin\phi_\bk\\
H_z(\bk)&=&\left|\mathbf{H}(\bk)\right|\cos\theta_\bk
\ee
with
\be
\left|\mathbf{H}(\bk)\right|&=&\sqrt{H_x(\bk)^2+H_y(\bk)^2+H_z(\bk)^2}.
\ee
The energy bands take the form
\be
E^\pm(\bk)&=&H_0(\bk)\pm\left|\mathbf{H}(\bk)\right|
\ee
and the coefficients $\{A^\pm,B^\pm\}$ are given by the corresponding eigenvectors. In order to consistently describe the phase across the whole sphere, we use a different gauge convention for the upper and lower hemispheres. For the northern hemisphere including $\theta_\bk=0$ we choose
\be
\left(\renewcommand\arraystretch{1.5}\begin{array}{c}
A^+_N\\
B^+_N\end{array}\right)&=&\left(\renewcommand\arraystretch{1.5}\begin{array}{c}
\cos\frac{\theta_\bk}{2}\\
\sin\frac{\theta_\bk}{2}e^{i\phi_\bk}\\
\end{array}\right)\\
\left(\renewcommand\arraystretch{1.5}\begin{array}{c}
A^-_N\\
B^-_N\end{array}\right)&=&\left(\renewcommand\arraystretch{1.5}\begin{array}{c}
-\sin\frac{\theta_\bk}{2}e^{-i\phi_\bk}\\
\cos\frac{\theta_\bk}{2}
\end{array}\right),
\ee
whilst for the southern hemisphere including $\theta_\bk=\pi$ we choose 
\be
\left(\renewcommand\arraystretch{1.5}\begin{array}{c}
A^+_S\\
B^+_S\end{array}\right)&=&\left(\renewcommand\arraystretch{1.5}\begin{array}{c}
\cos\frac{\theta_\bk}{2}e^{-i\phi_\bk}\\
\sin\frac{\theta_\bk}{2}\\
\end{array}\right)\\
\left(\renewcommand\arraystretch{1.5}\begin{array}{c}
A^-_S\\
B^-_S\end{array}\right)&=&\left(\renewcommand\arraystretch{1.5}\begin{array}{c}
-\sin\frac{\theta_\bk}{2}\\
\cos\frac{\theta_\bk}{2}e^{i\phi_\bk}
\end{array}\right),
\ee
We can obtain the southern wavefunction from the northern wavefunction by applying the $\mathbf{k}$-dependent gauge transformation $\{A^\pm_S,B^\pm_S\}=e^{\mp i\phi_\bk}\{A^\pm_N,B^\pm_N\}$.

We explicitly calculate the Berry connections as defined in Eq.~\ref{connection} with the wavefunctions as in Eq.~\ref{bloch} for the two hemisphere gauges, and find
\[
\renewcommand\arraystretch{1.5}
\begin{array}{ccccc}
\mathbf{A}_N^{--}&=&-\sin^2\frac{\theta_\bk}{2}\phi_\bk'+\cos^2\frac{\theta_\bk}{2}\ba&\to&-\sin^2\frac{\theta_\bk}{2}\phi_\bk'\\
\mathbf{A}_N^{++}&=&\sin^2\frac{\theta_\bk}{2}\phi_\bk'+\sin^2\frac{\theta_\bk}{2}\ba&\to&\sin^2\frac{\theta_\bk}{2}\phi_\bk'\\
\mathbf{A}_N^{-+}&=&\frac{1}{2}e^{i\phi_\bk}\left[\sin\theta_\bk\ba+\sin\theta_\bk\phi_\bk'-i\theta_\bk'\right]&\to&\frac{1}{2}e^{i\phi_\bk}\left[\sin\theta_\bk\phi_\bk'-i\theta_\bk'\right]\\
\mathbf{A}_N^{+-}&=&\frac{1}{2}e^{-i\phi_\bk}\left[\sin\theta_\bk\ba+\sin\theta_\bk\phi_\bk'+i\theta_\bk'\right]&\to&\frac{1}{2}e^{-i\phi_\bk}\left[\sin\theta_\bk\phi_\bk'+i\theta_\bk'\right]\\
\end{array}
\]
\[
\renewcommand\arraystretch{1.5}
\begin{array}{ccccc}
\mathbf{A}_S^{--}&=&\cos^2\frac{\theta_\bk}{2}\phi_\bk'+\cos^2\frac{\theta_\bk}{2}\ba&\to&\cos^2\frac{\theta_\bk}{2}\phi_\bk'\\
\mathbf{A}_S^{++}&=&-\cos^2\frac{\theta_\bk}{2}\phi_\bk'+\sin^2\frac{\theta_\bk}{2}\ba&\to&-\cos^2\frac{\theta_\bk}{2}\phi_\bk'\\
\mathbf{A}_S^{-+}&=&\frac{1}{2}e^{-i\phi_\bk}\left[\sin\theta_\bk\ba+\sin\theta_\bk\phi_\bk'-i\theta_\bk'\right]&\to&\frac{1}{2}e^{-i\phi_\bk}\left[\sin\theta_\bk\phi_\bk'-i\theta_\bk'\right]\\
\mathbf{A}_S^{+-}&=&\frac{1}{2}e^{i\phi_\bk}\left[\sin\theta_\bk\ba+\sin\theta_\bk\phi_\bk'+i\theta_\bk'\right]&\to&\frac{1}{2}e^{i\phi_\bk}\left[\sin\theta_\bk\phi_\bk'+i\theta_\bk'\right]\\
\end{array}
\]
where the prime indicates the gradient with respect to $\bk$, and where in the final column we have set the site displacement $\ba\to\mathbf{0}$ so that the sublattices overlap. We note that these connections satisfy $\mathbf{A}^{ab}=\left[\mathbf{A}^{ba}\right]^*$ and that the expressions for different hemispheres are related by the phase $e^{2i\phi_\bk}$.

We now calculate the Berry curvature from these connections using Eq.~\ref{curv},
\be
F^{--}_N&=&\frac{1}{2}\sin\theta_\bk\left[\nabla_\bk\phi_\bk\times\nabla_\bk\theta_\bk+\ba\times\nabla_\bk\theta_\bk\right]\\
&\to&\frac{1}{2}\sin\theta_\bk\left[\nabla_\bk\phi_\bk\times\nabla_\bk\theta_\bk\right]\\
F^{++}_N&=&-\frac{1}{2}\sin\theta_\bk\left[\nabla_\bk\phi_\bk\times\nabla_\bk\theta_\bk+\ba\times\nabla_\bk\theta_\bk\right]\\
&\to&-\frac{1}{2}\sin\theta_\bk\left[\nabla_\bk\phi_\bk\times\nabla_\bk\theta_\bk\right]\\
F^{-+}_N&=&\frac{1}{2}e^{i\phi_\bk}\left[-i\sin\theta_\bk\left(\ba\times\nabla_\bk\phi_\bk\right)-\cos\theta_\bk\left(\ba\times\nabla_\bk\theta_\bk\right)+2\sin^2\frac{\theta_\bk}{2}\left(\nabla_\bk\phi_\bk\times\nabla_\bk\theta_\bk\right)\right]\\
&\to&e^{i\phi_\bk}\sin^2\frac{\theta_\bk}{2}\left(\nabla_\bk\phi_\bk\times\nabla_\bk\theta_\bk\right)\\
F^{+-}_N&=&\frac{1}{2}e^{-i\phi_\bk}\left[i\sin\theta_\bk\left(\ba\times\nabla_\bk\phi_\bk\right)-\cos\theta_\bk\left(\ba\times\nabla_\bk\theta_\bk\right)+2\sin^2\frac{\theta_\bk}{2}\left(\nabla_\bk\phi_\bk\times\nabla_\bk\theta_\bk\right)\right]\\
&\to&e^{-i\phi_\bk}\sin^2\frac{\theta_\bk}{2}\left(\nabla_\bk\phi_\bk\times\nabla_\bk\theta_\bk\right)\\
\ee
\be
F^{--}_S&=&F^{--}_N\\
F^{++}_S&=&F^{++}_N\\
F^{-+}_S&=&\frac{1}{2}e^{-i\phi_\bk}\left[i\sin\theta_\bk\left(\ba\times\nabla_\bk\phi_\bk\right)-\cos\theta_\bk\left(\ba\times\nabla_\bk\theta_\bk\right)-2\cos^2\frac{\theta_\bk}{2}\left(\nabla_\bk\phi_\bk\times\nabla_\bk\theta_\bk\right)\right]\\
&\to&-e^{-i\phi_\bk}\cos^2\frac{\theta_\bk}{2}\left(\nabla_\bk\phi_\bk\times\nabla_\bk\theta_\bk\right)\\
F^{+-}_S&=&\frac{1}{2}e^{i\phi_\bk}\left[-i\sin\theta_\bk\left(\ba\times\nabla_\bk\phi_\bk\right)-\cos\theta_\bk\left(\ba\times\nabla_\bk\theta_\bk\right)-2\cos^2\frac{\theta_\bk}{2}\left(\nabla_\bk\phi_\bk\times\nabla_\bk\theta_\bk\right)\right]\\
&\to&-e^{i\phi_\bk}\cos^2\frac{\theta_\bk}{2}\left(\nabla_\bk\phi_\bk\times\nabla_\bk\theta_\bk\right)\\
\ee
In the final line of each calculation we have again set $\ba\to\mathbf{0}$ for simplicity. We see that for the off-diagonal terms $F^{ab}_S= e^{i\omega}F_N^{ab}+ie^{i\omega}\bigg[\boldsymbol{\partial}\omega\times\mathbf{A}^{ab}_N\bigg]$ with $\omega=\pm2\phi_\bk$.
\subsubsection{Nearest Neighbour Interaction}
To generate a state with non-zero expectation values $\langle c_{\pm,\bk}^\dagger c^{\phantom{\dagger}}_{\mp,\bk} \rangle$ we will switch on a weak nearest neighbour interaction, which we derive below.

We recall that the Haldane model may be obtained from the tight-binding model,
\be
\hat{H}&=&t_1\sum_{\bR\in A}\sum_i\left(a_\bR^\dagger b^{\phantom{\dagger}}_{\bR+\ba_i}+h.c.\right)\\
&&+t_2e^{i\phi}\sum_{\bR\in A}\left(a_{\bR+\bb_1}^\dagger a^{\phantom{\dagger}}_\bR+a_{\bR+\bb_1+\bb_3}^\dagger a^{\phantom{\dagger}}_{\bR+\bb_1}+a_\bR^\dagger a^{\phantom{\dagger}}_{\bR+\bb_1+\bb_3}\right)+h.c.\\
&&+(a\to b,\phi\to-\phi)
\ee
through the Fourier transform
\be
a_\bR&=&\frac{1}{\sqrt{\mathcal{A}}}\sum_{\bk}a_\bk e^{-i\bk\cdot\bR}\\
a^\dagger_\bR&=&\frac{1}{\sqrt{\mathcal{A}}}\sum_{\bk}a^\dagger_\bk e^{i\bk\cdot\bR}\\
b_\bR&=&\frac{1}{\sqrt{\mathcal{A}}}\sum_{\bk}b_\bk e^{-i\bk\cdot\bR}\\
b^\dagger_\bR&=&\frac{1}{\sqrt{\mathcal{A}}}\sum_{\bk}b^\dagger_\bk e^{i\bk\cdot\bR}\\
\ee
etc. Here we have introduced the operators $\{a^\dagger_\br,b^\dagger_\br\}$ which create states on the $\{A,B\}$ sublattice site in the unit cell defined by $\bR$. $\mathcal{A}$ is the system area factor needed to preserve the anticommutation relations. The Fourier operators may be identified with the orbital band creation operators from earlier
\be
\{a^\dagger_\bk,b^\dagger_\bk\}\leftrightarrow c^\dagger_{\{A,B\},\bk}.
\ee

We introduce a nearest neighbour (density-density) interaction term,
\be
\hat{U}&=&-U\sum_{\bR\in A}\hat{n}_{\bR}\left(\hat{n}_{\bR+\ba_1}+\hat{n}_{\bR+\ba_2}+\hat{n}_{\bR+\ba_3}\right)+h.c.\\
&=&-U\sum_{\bR\in A}a^\dagger_\bR a_\bR\left(b^\dagger_{\bR+\ba_1} b_{\bR+\ba_1}+b^\dagger_{\bR+\ba_2} b_{\bR+\ba_2}+b^\dagger_{\bR+\ba_3} b_{\bR+\ba_3}\right)+h.c.\\
&=&U\sum_{\bR\in A}\left(a^\dagger_\bR b^\dagger_{\bR+\ba_1} a_\bR b_{\bR+\ba_1}+a^\dagger_\bR b^\dagger_{\bR+\ba_2}a_\bR  b_{\bR+\ba_2}+a^\dagger_\bR b^\dagger_{\bR+\ba_3}a_\bR  b_{\bR+\ba_3}\right)+h.c.
\ee
where in the final line we have normal ordered. Taking the Fourier transform we find
\be
U(\bk_1,\bk_2,\bk_3,\bk_4)&=&\frac{U}{\mathcal{A}^2}\sum_{\br\in A}\sum_{\bk_1\bk_2\bk_3\bk_4}\sum_je^{i\br\cdot(\bk_1+\bk_2-\bk_3-\bk_4)}e^{i\ba_j\cdot(\bk_2-\bk_4)}a^\dagger_{\bk_1} b^\dagger_{\bk_2} a_{\bk_3} b_{\bk_4}+h.c.\\
&=&\frac{U}{\mathcal{A}}\sum_{\bk_1\bk_2\bk_3\bk_4}\delta_{\bk_1+\bk_2,\bk_3+\bk_4}\sum_je^{i\ba_j\cdot(\bk_2-\bk_4)}a^\dagger_{\bk_1} b^\dagger_{\bk_2} a_{\bk_3} b_{\bk_4}+h.c.
\ee

To convert to the band basis from the orbital basis, we use the eigenvectors from earlier to write
\be
\left(\begin{array}{c}
c^\dagger_{+\bk}\\
c^\dagger_{-\bk}
\end{array}\right)&=&\left(\begin{array}{cc}
A^+ & B^+ \\
A^- & B^- 
\end{array}\right)\left(\begin{array}{c}
a^\dagger_\bk\\
b^\dagger_\bk
\end{array}\right)
\ee
and so
\be
\left(\begin{array}{c}
a^\dagger_{\bk}\\
b^\dagger_{\bk}
\end{array}\right)&=&\left(\begin{array}{cc}
\left(A^+\right)^* & \left(A^-\right)^* \\
\left(B^+\right)^* & \left(B^-\right)^* 
\end{array}\right)\left(\begin{array}{c}
c^\dagger_{+\bk}\\
c^\dagger_{-\bk}
\end{array}\right),
\ee
where the values of $\{A^\pm(\bk),B^\pm(\bk)\}$ were given previously.

This allows us to write out the interaction in terms of the band creation and annihilation operators,
\be
\hat{U}(\bk_1,\bk_2,\bk_3,\bk_4)&=&\frac{U}{\mathcal{A}}\sum_{\bk_1\bk_2\bk_3\bk_4}\delta_{\bk_1+\bk_2,\bk_3+\bk_4}\sum_je^{i\ba_j\cdot(\bk_2-\bk_4)}\left(\left[A^+(\bk_1)\right]^*c^\dagger_{+\bk_1}+\left[A^-(\bk_1)\right]^*c^\dagger_{-\bk_1}\right)\times\\
&& \left(\left[B^+(\bk_2)\right]^*c^\dagger_{+\bk_2}+\left[B^-(\bk_2)\right]^*c^\dagger_{-\bk_2}\right) \left(\left[A^+(\bk_3)\right]c^{\phantom{\dagger}}_{+\bk_3}+\left[A^-(\bk_3)\right]c^{\phantom{\dagger}}_{-\bk_3}\right)\times\\
&& \left(\left[B^+(\bk_4)\right]c^{\phantom{\dagger}}_{+\bk_4}+\left[B^-(\bk_4)\right]c^{\phantom{\dagger}}_{-\bk_4}\right)\delta_{\bp+\bq,\br+\bs}+h.c.\\
&\equiv&\sum_{\bp,\bq,\br,\bs}\sum_{\alpha,\beta,\gamma,\delta}U_{\alpha\beta\gamma\delta}(\bp,\bq,\br,\bs)c^\dagger_{\alpha\mathbf{p}} c^\dagger_{\beta\mathbf{q}} c_{\gamma\mathbf{r}} c_{\delta\mathbf{s}} \delta_{\mathbf{p}+\mathbf{q},\mathbf{r}+\mathbf{s}} +h.c.
\ee
where the greek indices may take either sign $\pm$. We can pick out the function $U$ as
\be
U_{\alpha\beta\gamma\delta}(\bp,\bq,\br,\bs)&=&\frac{U}{\mathcal{A}}\sum_je^{i\ba_j\cdot(\bq-\bs)}\left[A^\alpha(\bp)\right]^*\left[B^\beta(\bq)\right]^*\left[A^\gamma(\br)\right]\left[B^\delta(\bs)\right].
\ee

However, the interaction is antisymmetric under the exchange of $(\alpha\bp\leftrightarrow\beta\bq)$ or $(\gamma\br\leftrightarrow\delta\bs)$, and it is useful to transfer this antisymmetry to our expression for $U$ above. To find the correctly antisymmetrised $u_{\alpha\beta\gamma\delta}(\bp,\bq,\br,\bs)$ we calculate
\be
u_{\alpha\beta\gamma\delta}(\bp,\bq,\br,\bs)&=&\bra{0}c_{\alpha\bp}c_{\beta\bq}\hat{U}c^\dagger_{\gamma\br}c^\dagger_{\delta\bs}\ket{0}\\
&=&U_{\alpha\beta\gamma\delta}(\bp,\bq,\br,\bs)-U_{\beta\alpha\gamma\delta}(\bq,\bp,\br,\bs)-U_{\alpha\beta\delta\gamma}(\bp,\bq,\bs,\br)+U_{\beta\alpha\delta\gamma}(\bq,\bp,\bs,\br)
\ee

This allows us to write the interaction as 
\be
\hat{U}(\bk_1,\bk_2,\bk_3,\bk_4)&=&\frac{1}{4}\sum_{\bp,\bq,\br,\bs}\sum_{\alpha,\beta,\gamma,\delta}u_{\alpha\beta\gamma\delta}(\bp,\bq,\br,\bs)c^\dagger_{\alpha\mathbf{p}} c^\dagger_{\beta\mathbf{q}} c_{\gamma\mathbf{r}} c_{\delta\mathbf{s}} \delta_{\mathbf{p}+\mathbf{q},\mathbf{r}+\mathbf{s}} +h.c.
\ee
but where $u_{\alpha\beta\gamma\delta}(\bp,\bq,\br,\bs)$ is now antisymmeric as required. We can include the effect of the Hermitian conjugate to find (after relabelling)
\be
\hat{U}(\bk_1,\bk_2,\bk_3,\bk_4)&=&\frac{1}{4}\sum_{\bp,\bq,\br,\bs}\sum_{\alpha,\beta,\gamma,\delta}\left[u_{\alpha\beta\gamma\delta}(\bp,\bq,\br,\bs)+u^*_{\delta\gamma\beta\alpha}(\bs,\br,\bq,\bp)\right]c^\dagger_{\alpha\mathbf{p}} c^\dagger_{\beta\mathbf{q}} c_{\gamma\mathbf{r}} c_{\delta\mathbf{s}} \delta_{\mathbf{p}+\mathbf{q},\mathbf{r}+\mathbf{s}}.
\ee
\subsection{Perturbation Theory}
We now consider this interaction $\lambda\hat{U}$ as a perturbation to a simple many-body ground state, and calculate the resulting off-diagonal occupation numbers $\bar{n}^{\pm\mp}_\bk=\langle c_{\pm,\bk}^\dagger c^{\phantom{\dagger}}_{\mp,\bk} \rangle$. 

We define the unperturbed ground state of the system to have the lower band completely filled and the upper band completely empty,
\be
\ket{\mathrm{Ground}}&=&\prod_{\bk\in\mathrm{BZ}}c^{\dagger}_{-\bk}\ket{0}.
\ee
If we apply the perturbation $\lambda\hat{U}$, then the perturbed many-body wavefunction may be written to first order as
\be
\ket{\widetilde{\mathrm{Ground}}}&=&\ket{\mathrm{Ground}}+\lambda\sum_{E\neq\mathrm{Ground}}\ket{E}\frac{\bra{E}\hat{U}\ket{\mathrm{Ground}}}{E_0-E_E}.
\ee

To begin we consider a simple case where the interaction just depends on four specific momenta and band labels,
\be
\hat{U}&=&c^\dagger_{a\mathbf{p}} c^\dagger_{b\mathbf{q}} c_{c\mathbf{r}} c_{d\mathbf{s}} \delta_{\mathbf{p}+\mathbf{q},\mathbf{r}+\mathbf{s}},
\ee
which upon acting on the ground state leads to
\be
\hat{U}\ket{\mathrm{Ground}}&=&\delta_{c-}\delta_{d-}\bigg[\delta_{a+}\delta_{b+}(-1)^{P(\bp,\bq,\br,\bs)}\ket{n_\bp^+=1,n_\bq^+=1;n_\br^-=0,n_\bs^-=0}\\
&&+\delta_{a+}\delta_{b-}\delta_{\bq\bs}(-1)^{P(\bp,\bq,\br,\bq)}\ket{n_\bp^+=1;n_\br^-=0}\\
&&+\delta_{a+}\delta_{b-}\delta_{\bq\br}(-1)^{P(\bp,\bq,\bq,\bs)}\ket{n_\bp^+=1;n_\bs^-=0}\\
&&+\delta_{a-}\delta_{b+}\delta_{\bp\bs}(-1)^{P(\bp,\bq,\br,\bp)}\ket{n_\bq^+=1;n_\br^-=0}\\
&&+\delta_{a-}\delta_{b+}\delta_{\bp\br}(-1)^{P(\bp,\bq,\bp,\bs)}\ket{n_\bq^+=1;n_\bs^-=0}\\
&&+\delta_{a-}\delta_{b-}\delta_{\bp\br}\delta_{\bq\bs}(-1)^{P(\bp,\bq,\bp,\bq)}\ket{\mathrm{Ground}}\\
&&+\delta_{a-}\delta_{b-}\delta_{\bp\bs}\delta_{\bq\br}(-1)^{P(\bp,\bq,\bq,\bp)}\ket{\mathrm{Ground}}\bigg]
\ee
Here we have only indicated the states in the upper band that are occupied and the states in the lower band that are unoccupied, and we have introduced $(-1)^P$ as a sign permutation factor that comes from anticommuting the fermion operators. 

This corresponds to the first order wavefunction correction ($\{-,+\}\leftrightarrow\{0,1\}$)
\be
\ket{\widetilde{\mathrm{Ground}}}&=&\ket{\mathrm{Ground}}-\frac{\lambda}{2\Delta}\delta_{c0}\delta_{d0}\delta_{a1}\delta_{b1}(-1)^{P(\bp,\bq,\br,\bs)}\ket{n_\bp^1=1,n_\bq^1=1;n_\br^0=0,n_\bs^0=0}\\
&&-\frac{\lambda}{\Delta}\delta_{c0}\delta_{d0}\bigg[\delta_{a1}\delta_{b0}\delta_{\bq\bs}(-1)^{P(\bp,\bq,\br,\bq)}\ket{n_\bp^1=1;n_\br^0=0}\\
&&+\delta_{a1}\delta_{b0}\delta_{\bq\br}(-1)^{P(\bp,\bq,\bq,\bs)}\ket{n_\bp^1=1;n_\bs^0=0}\\
&&+\delta_{a0}\delta_{b1}\delta_{\bp\bs}(-1)^{P(\bp,\bq,\br,\bp)}\ket{n_\bq^1=1;n_\br^0=0}\\
&&+\delta_{a0}\delta_{b1}\delta_{\bp\br}(-1)^{P(\bp,\bq,\bp,\bs)}\ket{n_\bq^1=1;n_\bs^0=0}\bigg]
\ee
where we have further assumed that the bands are flat and have an energy gap of $\Delta$ (i.e. the bands have been flattened by a local, single-particle term in the Hamiltonian that does not change the single-particle wavefunctions).

The nearest neighbour interaction discussed previously can be written
\be
\hat{U} = \sum_{\mathbf{p},\mathbf{q},\mathbf{r},\mathbf{s}}\sum_{\alpha,\beta,\gamma,\delta} u_{\alpha\beta\gamma\delta}(\mathbf{p},\mathbf{q},\mathbf{r},\mathbf{s}) c^\dagger_{\alpha\mathbf{p}} c^\dagger_{\beta\mathbf{q}} c_{\gamma\mathbf{r}} c_{\delta\mathbf{s}} \delta_{\mathbf{p}+\mathbf{q},\mathbf{r}+\mathbf{s}}
\ee
where we have relabelled
\be
\frac{1}{4}\left[u_{\alpha\beta\gamma\delta}(\bp,\bq,\br,\bs)+u^*_{\delta\gamma\beta\alpha}(\bs,\br,\bq,\bp)\right]\to u_{\alpha\beta\gamma\delta}(\bp,\bq,\br,\bs).
\ee
This leads to the first order wavefunction
\be
\ket{\widetilde{\rm Ground}}&=&\ket{\rm Ground}+\lambda\sum_{E\neq\mathrm{Ground}}\ket{E}\frac{\bra{E}\hat{U}\ket{\mathrm{Ground}}}{E_0-E_E}\\
&=&\ket{\rm Ground}\\
&&-\frac{\lambda}{2\Delta}\sum_{\mathbf{p},\mathbf{q},\mathbf{r},\mathbf{s}}u_{1100}(\mathbf{p},\mathbf{q},\mathbf{r},\mathbf{s})\left(-1\right)^{P(\bp,\bq,\br,\bs)}\ket{n^1_\bp=1, n^1_\bq=1;n^0_\br=0,n^0_\bs=0}\\
&&-\frac{\lambda}{\Delta}\sum_{\bp\bq\br}u_{1000}(\bp,\bq,\br,\bq)\left(-1\right)^{P(\bp,\bq,\br,\bq)}\ket{n^1_\bp=1;n^0_\br=0}\\
&&-\frac{\lambda}{\Delta}\sum_{\bp\bq\bs}u_{1000}(\bp,\bq,\bq,\bs)\left(-1\right)^{P(\bp,\bq,\bq,\bs)}\ket{n^1_\bp=1;n^0_\bs=0}\\
&&-\frac{\lambda}{\Delta}\sum_{\bp\bq\br}u_{0100}(\bp,\bq,\br,\bp)\left(-1\right)^{P(\bp,\bq,\br,\bp)}\ket{n^1_\bq=1;n^0_\br=0}\\
&&-\frac{\lambda}{\Delta}\sum_{\bp\bq\bs}u_{0100}(\bp,\bq,\bp,\bs)\left(-1\right)^{P(\bp,\bq,\bp,\bs)}\ket{n^1_\bq=1;n^0_\bs=0}.
\ee

We are interested in the two off-diagonal expectation values $\bar{n}^{\pm\mp}_\bk=\langle c^\dagger_{\pm\bk}c_{\mp\bk}\rangle$, whose operators transfer a single electron from one band to the other. For this reason, there will be no contribution from the first term in the perturbation at first order.

For the ordering convention, we will assume a general state takes the form
\be
\ket{\mathrm{State}}&=&c^\dagger_{-\bk_1}c^\dagger_{+\bk_1}c^\dagger_{-\bk_2}c^\dagger_{+\bk_2}\ldots c^\dagger_{-\bk_N}c^\dagger_{+\bk_N}\ket{0}
\ee
where operators are ordered (from the left) first by momentum and then by band. We assume that there are $N$ values of momentum that can be ordered consistently for any $N$. A general state simply has operators missing from the above definition. 

We find that the relevant contributions to $\bar{n}^{+-}_\bk$ are
\be
c^\dagger_{+\bk}c^{\phantom{\dagger}}_{-\bk}\ket{\widetilde{\rm Ground}}&=&\ket{n^1_\bk=1;n^0_\bk=0}+O(\lambda)
\ee
where the first order terms are not proportional to $\ket{\rm Ground}$ and so will not contribute at first order. Next,
\be
\bra{\widetilde{\rm Ground}}c^\dagger_{+\bk}c^{\phantom{\dagger}}_{-\bk}\ket{\widetilde{\rm Ground}}&=&-\frac{\lambda}{\Delta}\left[\sum_{\bp\bq\br}u^*_{1000}(\bp,\bq,\br,\bq)\left(-1\right)^{P(\bp,\bq,\br,\bq)}\delta_{\bp \bk}\delta_{\br\bk}\right.\\
&&+\sum_{\bp\bq\bs}u^*_{1000}(\bp,\bq,\bq,\bs)\left(-1\right)^{P(\bp,\bq,\bq,\bs)}\delta_{\bp \bk}\delta_{\bs\bk}\\
&&+\sum_{\bp\bq\br}u^*_{0100}(\bp,\bq,\br,\bp)\left(-1\right)^{P(\bp,\bq,\br,\bp)}\delta_{\bq \bk}\delta_{\br\bk}\\
&&\left.+\sum_{\bp\bq\bs}u^*_{0100}(\bp,\bq,\bp,\bs)\left(-1\right)^{P(\bp,\bq,\bp,\bs)}\delta_{\bq \bk}\delta_{\bs\bk}\right]
\ee
\be
&=&-\frac{\lambda}{\Delta}\left[\sum_{\bq}u^*_{1000}(\bk,\bq,\bk,\bq)\left(-1\right)^{P(\bk,\bq,\bk,\bq)}+\sum_{\bq}u^*_{1000}(\bk,\bq,\bq,\bk)\left(-1\right)^{P(\bk,\bq,\bq,\bk)}\right.\\
&&\left.+\sum_{\bp}u^*_{0100}(\bp,\bk,\bk,\bp)\left(-1\right)^{P(\bp,\bk,\bk,\bp)}+\sum_{\bp}u^*_{0100}(\bp,\bk,\bp,\bk)\left(-1\right)^{P(\bp,\bk,\bp,\bk)}\right].
\ee
Using the antisymmetry of the sign factor $(-1)^P$ and the antisymmetry of the function $u$ under the interchange of the first or last pair of quantum numbers, these four terms can be combined,
\be
\bar{n}^{+-}_\bk&=&-\frac{4\lambda}{\Delta}\sum_{\bq}u^*_{1000}(\bk,\bq,\bk,\bq)\left(-1\right)^{P(\bk,\bq,\bk,\bq)}
\ee
Finally, we note that the sign factor $\left(-1\right)^{P(\bk,\bq,\bk,\bq)}$ comes from anti commuting the operator $c^\dagger_{1\bk}c^\dagger_{0\bq}c^{\phantom{\dagger}}_{0\bk}c^{\phantom{\dagger}}_{0\bq}$ through the operators in the definition of $\ket{\rm Ground}$ according to our sign convention. We find that
\be
c^\dagger_{1\bk}c^\dagger_{0\bq}c^{\phantom{\dagger}}_{0\bk}c^{\phantom{\dagger}}_{0\bq}\ket{\rm Ground}&=&-c^\dagger_{1\bk}c^{\phantom{\dagger}}_{0\bk}c^\dagger_{0\bq}c^{\phantom{\dagger}}_{0\bq}\ket{\rm Ground}\\
&=&-\ket{n_\bk^1=1;n_\bk^0=0}
\ee
and so $\left(-1\right)^{P(\bk,\bq,\bk,\bq)}=-1$ and 
\be
\bar{n}^{+-}_\bk&=&\frac{4\lambda}{\Delta}\sum_{\bq}u^*_{1000}(\bk,\bq,\bk,\bq).
\ee
We find similarly that
\be
\bar{n}^{-+}_\bk=\bra{\widetilde{\rm Ground}}c^\dagger_{-\bk}c^{\phantom{\dagger}}_{+\bk}\ket{\widetilde{\rm Ground}}&=&\frac{4\lambda}{\Delta}\sum_\bq u_{1000}(\bk,\bq,\bk,\bq),
\ee
and as required, $\bar{n}^{-+}_\bk=\left[\bar{n}^{+-}_\bk\right]^*$.
\subsection{Expression for $\bv$}
We now calculate the explicit value for $\bar{n}^{-+}_\bk$ by substituting in for the nearest neighbour interaction considered previously.
\be
\bar{n}^{-+}_\bk&=&\frac{4\lambda}{\Delta}\sum_\bq u_{1000}(\bk,\bq,\bk,\bq)\\
&\to&\frac{4\lambda}{\Delta}\sum_\bq\frac{1}{4}\left[u_{1000}(\bk,\bq,\bk,\bq)+u^*_{0001}(\bq,\bk,\bq,\bk)\right]\\
&=&\frac{\lambda}{\Delta}\sum_\bq\left[U_{1000}(\bk,\bq,\bk,\bq)-U_{0100}(\bq,\bk,\bk,\bq)-U_{1000}(\bk,\bq,\bq,\bk)+U_{0100}(\bq,\bk,\bq,\bk)\right.\\
&&\left.+U^*_{0001}(\bq,\bk,\bq,\bk)-U^*_{0001}(\bk,\bq,\bq,\bk)-U^*_{0010}(\bq,\bk,\bk,\bq)+U^*_{0010}(\bk,\bq,\bk,\bq)\right]\\
\ee
\be
&=&\frac{2U\lambda}{\Delta\mathcal{A}}\sum_\bq\left[\sum_j\left[A^+(\bk)\right]^*\left[B^-(\bq)\right]^*\left[A^-(\bk)\right]\left[B^-(\bq)\right]-\sum_je^{i\ba_j\cdot(\bk-\bq)}\left[A^-(\bq)\right]^*\left[B^+(\bk)\right]^*\left[A^-(\bk)\right]\left[B^-(\bq)\right]\right.\\
&&\left.-\sum_je^{i\ba_j\cdot(\bq-\bk)}\left[A^+(\bk)\right]^*\left[B^-(\bq)\right]^*\left[A^-(\bq)\right]\left[B^-(\bk)\right]+\sum_j\left[A^-(\bq)\right]^*\left[B^+(\bk)\right]^*\left[A^-(\bq)\right]\left[B^-(\bk)\right]\right].
\ee
Here we have assumed that the interaction strength $U$ is real.

If we choose the northern hemisphere gauge we find
\be
\bar{n}^{-+}_{N,\bk}&=&\frac{U\lambda}{2\Delta\mathcal{A}}e^{-i\phi_{\bk}}\sum_\bq\sum_j\left[-\sin{\theta_\bk}\left(1+\cos\theta_\bq\right)\right.\\
&&-e^{i\ba_j\cdot(\bk-\bq)} e^{i\left(\phi_\bq-\phi_\bk\right)}\left(1-\cos\theta_\bk\right)\sin{\theta_\bq}\\
&&+e^{i\ba_j\cdot(\bq-\bk)}e^{-i\left(\phi_{\bq}-\phi_\bk\right)}\left(1+\cos\theta_\bk\right)\sin{\theta_\bq}\\
&&\left.+\sin{\theta_\bk}\left(1-\cos\theta_\bq\right)\right].
\ee
We now define 
\be
r_{\bq\bk}e^{i\varphi_{\bq\bk}}=\sum_je^{i\ba_j\cdot\left(\bq-\bk\right)}
\ee
or equivalently
\be
\varphi_{\bq\bk}&\equiv&\phi_{\bq-\bk}\\
r_{\bq\bk}&\equiv&(1/t_1)\left|\mathbf{H}(\bq-\bk)\right|\sin\theta_{\bq-\bk}.
\ee
so that
\be
\bar{n}^{-+}_{N,\bk}&=&\frac{U\lambda}{\Delta\mathcal{A}}e^{-i\phi_{\bk}}\sum_\bq\left[-3\sin{\theta_\bk}\cos\theta_\bq+r_{\bq\bk}\sin\theta_\bq\cos\theta_\bk\cos\left(\varphi_{\bq\bk}-\phi_\bq+\phi_\bk\right)\right.\\
&&\left.+ir_{\bq\bk}\sin\theta_\bq\cos\theta_\bk\sin\left(\varphi_{\bq\bk}-\phi_\bq+\phi_\bk\right)\right].
\ee
In the southern hemisphere we find
\be
\bar{n}^{-+}_{S,\bk}&=&\frac{U\lambda}{\Delta\mathcal{A}}e^{i\phi_{\bk}}\sum_\bq\left[-3\sin{\theta_\bk}\cos\theta_\bq+r_{\bq\bk}\sin\theta_\bq\cos\theta_\bk\cos\left(\varphi_{\bq\bk}-\phi_\bq+\phi_\bk\right)\right.\\
&&\left.+ir_{\bq\bk}\sin\theta_\bq\cos\theta_\bk\sin\left(\varphi_{\bq\bk}-\phi_\bq+\phi_\bk\right)\right],
\ee
which differs by a factor of $e^{2i\phi_\bk}$ as expected. 

Using either gauge, we obtain the final expression for $\bv$
\be
 {\bf v} = {\rm Im} \left[  {\bf A} ^{-+}  \langle c_{-,\bk}^\dagger c_{+,\bk} \rangle \right] &=& \frac{1}{2}\sin\theta_\bk\left(\nabla\phi_\bk\right)\mathrm{Im}\left[\bar{n}^{-+}_\bk\right]-\frac{1}{2}\left(\nabla\theta_\bk\right)\mathrm{Re}\left[\bar{n}^{-+}_\bk\right]\\
 &=&\frac{U\lambda}{\Delta\mathcal{A}}\sum_\bq\bigg\{\frac{1}{2}\sin\theta_\bk\left(\nabla\phi_\bk\right)\left[r_{\bq\bk}\sin\theta_\bq\cos\theta_\bk\sin\left(\varphi_{\bq\bk}-\phi_\bq+\phi_\bk\right)\right]\\
 &&-\frac{1}{2}\left(\nabla\theta_\bk\right)\left[-3\sin{\theta_\bk}\cos\theta_\bq+r_{\bq\bk}\sin\theta_\bq\cos\theta_\bk\cos\left(\varphi_{\bq\bk}-\phi_\bq+\phi_\bk\right)\right]\bigg\}.
\ee
\subsection{Evaluation of $\nabla\times\bv$}
We take the curl of the above expression with respect to $\bk$ and find
\be
\nabla_\bk\times\bv&=&\frac{U\lambda}{2\Delta\mathcal{A}}\sum_\bq\bigg\{\bigg[\cos\theta_\bk\big[r_{\bq\bk}\sin\theta_\bq\cos\theta_\bk\sin\left(\varphi_{\bq\bk}-\phi_\bq+\phi_\bk\right)\big](\nabla_\bk\theta_\bk)\\
&&+\sin\theta_\bk\big[\sin\theta_\bq\cos\theta_\bk\sin\left(\varphi_{\bq\bk}-\phi_\bq+\phi_\bk\right)\big](\nabla_\bk r_{\bq\bk})\\
&&-\sin\theta_\bk\big[r_{\bq\bk}\sin\theta_\bq\sin\theta_\bk\sin\left(\varphi_{\bq\bk}-\phi_\bq+\phi_\bk\right)\big](\nabla_\bk\theta_\bk)\\
&&+\sin\theta_\bk\big[r_{\bq\bk}\sin\theta_\bq\cos\theta_\bk\cos\left(\varphi_{\bq\bk}-\phi_\bq+\phi_\bk\right)\big]\big(\nabla_\bk\varphi_{\bq\bk}\big)\bigg]\times(\nabla_\bk\phi_\bk)\\
&&-\bigg[\sin\theta_\bq\cos\theta_\bk\cos\left(\varphi_{\bq\bk}-\phi_\bq+\phi_\bk\right)\left(\nabla_\bk r_{\bq\bk}\right)\\
&&-r_{\bq\bk}\sin\theta_\bq\cos\theta_\bk\sin\left(\varphi_{\bq\bk}-\phi_\bq+\phi_\bk\right)\big(\nabla_\bk\left(\varphi_{\bq\bk}+\phi_\bk\right)\big)\bigg]\times\left(\nabla_\bk\theta_\bk\right)\bigg\}.
\ee
where we have used $\nabla\times(\psi\nabla\phi)=(\nabla\psi)\times(\nabla\phi)$ and $\nabla_\bk\phi_\bk\times\nabla_\bk\phi_\bk=0$ etc. We are looking for a value of $\bk$ for which $\nabla_\bk\times\bv\neq0$. We note that $\nabla\times\bv$ is intensive due to the factor of the system size in the initial numerator. We will later take the thermodynamic limit and convert the sum over $\bq$ to an integral.

It turns out that the `nice' choices of $\bk$, for which $\cos\theta_\bk=0$ or $\sin\theta_\bk=0$, have $\nabla\times\bv=0$. We will instead calculate $\nabla\times\bv$ at intermediate points in the Brillouin zone, numerically evaluating the sum over $\bq$.

We will use the lattice vectors defined previously and choose the Haldane model parameters
\be
t_1&=&1\\
t_2&=&1/4\\
\phi&=&\pi/2\\
M&=&0\\
a&=&1.
\ee
We will also need two reciprocal lattice vectors $\bB_j$, which we find (derived from the real space lattice vectors $\bb_j$) are
\be
\bB_1&=&\frac{4\pi}{3a}\left(-\frac{\sqrt{3}}{2},\frac{1}{2}\right)\\
\bB_2&=&\frac{4\pi}{3a}\left(-\frac{\sqrt{3}}{2},-\frac{1}{2}\right).
\ee
When we carry out the integration over $\bq$, we will choose a rhomboidal Brillouin zone defined by these lattice vectors and comprising $L$ points in each direction. 

In the thermodynamic limit we write
\be
\frac{U\lambda}{2\Delta\mathcal{A}}\sum_\bq f_\bq\equiv\frac{U\lambda}{2\Delta\mathcal{A}\Delta_\bq}\sum_\bq f_\bq\Delta_\bq\to\frac{U\lambda}{2\Delta\mathcal{A}\Delta_\bq}\int\mathrm{d}^2\bq \,f(\bq),
\ee
where $\Delta_\bq$ is the area element for each point in $\bq$-space, which is inversely proportional to the system size $\mathcal{A}$. Taking $L$ unit cells in each direction (so that $\mathcal{A}\equiv L^2$), the area element is the area of the parallelogram
\be
\Delta_\bq\equiv\left|\frac{\bB_1\times\bB_2}{L^2}\right|&=&\frac{8\sqrt{3}\pi^2}{9a^2L^2},
\ee
and so the thermodynamic integral may be written
\be
\frac{9\sqrt{3}U\lambda}{48\pi^2}\int\mathrm{d}^2\bq\,f(\bq).
\ee
Numerically we will calculate the quantity $\int\mathrm{d}^2\bq\,f(\bq)$ by reverting to the discrete sum $\sum_\bq f_\bq\Delta_\bq$.

To begin we choose $\bk=(\pi/4,\pi/3)$ which appears to give a non-zero value for $\nabla_\bk\times\bv$. We substitute this value of $\bk$ into our expression for $\nabla\times\bv$ in Mathematica, and sum over an $L\times L$ lattice of $\bq$-points in the Brillouin zone.

We find, as we increase $L$ (which increases the fineness of the grid for the integration over $\bq$), that the value tends towards $0.613596$ (ignoring the prefactor $9\sqrt{3}U\lambda/(48\pi^2)$). This convergence is shown in Figure~\ref{curlconv}.

\begin{figure}[h]
\begin{center}
\includegraphics[scale=0.80]{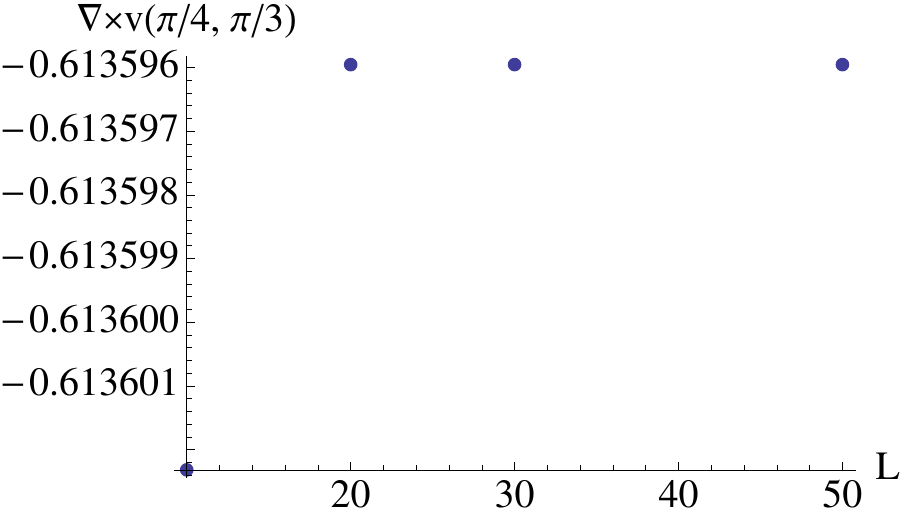}
\caption{Convergence towards value for $\nabla_\bk\times\bv$ with integration grid size $L\in\{10,20,30,50\}$, evaluated at $\bk=(\pi/4,\pi/3)$.\label{curlconv}}
\end{center}
\end{figure}

$L=20$ seems to give a good convergence to 5 decimal places.

We also choose $11\times11$ $\bk$-points across the Brillouin zone and calculate $\nabla\times\bv$ for each of them (with $L=20$). The results are plotted in Figure~\ref{curlv}. In particular, although the curl of $\bv$ vanishes along the boundary of the Brillouin zone, it appears to be non-zero in general in the centre. This structure repeats if we shift $\bk$ by a reciprocal lattice vector.

As mentioned above, if $\nabla\times\bv$ does not vanish throughout the Brillouin zone, we can choose a $\bk$-dependent gauge transformation that will cause Eq.~\ref{extraterm} to be non-zero, which in turn will cause $\sigma_{\rm NSCM}$ to be gauge dependent.

\begin{figure}[h]
\begin{center}
\includegraphics[scale=0.50]{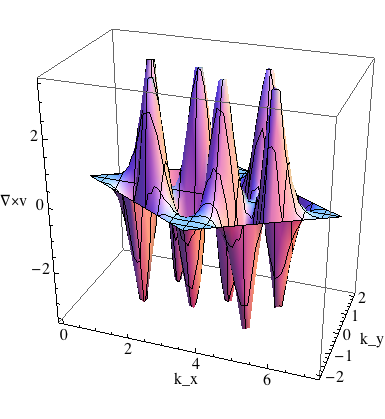}

\includegraphics[scale=0.50]{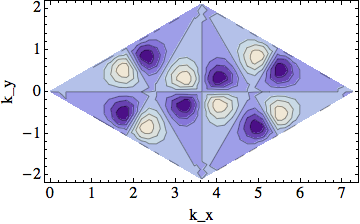}\hspace{1cm}
\includegraphics[scale=0.60]{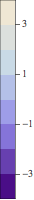}
\caption{$\nabla_\bk\times\bv$ evaluated at 121 points across the Brillouin zone for $L=20$.\label{curlv}}
\end{center}
\end{figure}

\end{widetext}
\end{document}